\documentclass[aps,prl,showpacs,showkeys,floats,reprint,superscriptaddress,floatfix,preprintnumbers]{revtex4-1}
\usepackage{amsfonts,amssymb,amsmath}
\usepackage{graphics,longtable,subfigure,epsfig,bm,ulem,color}
\graphicspath{{./fig/}{./png/}}
\def \xxi {\bm{\xi}}
\def \xizero {\bm{\xi}_{\rm 0}}

\urldef\ytubeurl\url{http://www.youtube.com/watch?v=ANDUNDHYgzk }{}

\begin{document}
\preprint{NORDITA-2011-39,NSF-KITP-11-140}
\title{Dynamic Multiscaling in Two-dimensional Fluid Turbulence}
\author{Samriddhi Sankar Ray} 
\email{samriddhisankarray@gmail.com}
\affiliation{Laboratoire Cassiop\'ee, Observatoire de la C\^ote 
d'Azur, UNS, CNRS, BP 4229, 06304 Nice Cedex 4, France}
\author{Dhrubaditya Mitra} 
\email{dhruba.mitra@gmail.com}
\affiliation{NORDITA, Roslagstullsbacken 23, SE-10691 Stockholm, Sweden }
\altaffiliation[visitor~]{Kavli Institute for Theoretical Physics, Santa Barbara.}
\author{Prasad Perlekar}
\email{p.perlekar@tue.nl}
\affiliation{Department of Mathematics and Computer Science,
Eindhoven University of Technology, P.O. Box 513, 5600 MB Eindhoven,
The Netherlands} 
\author{Rahul Pandit}
\email{rahul@physics.iisc.ernet.in}
\altaffiliation[\\ also at~]{Jawaharlal Nehru Centre For Advanced
Scientific Research, Jakkur, Bangalore, India}
\affiliation{Centre for Condensed Matter Theory, Department of Physics, Indian
Institute of Science, Bangalore 560012, India} 
\begin{abstract}
We obtain, by extensive direct numerical simulations, time-dependent and
equal-time structure functions for the vorticity, in both
quasi-Lagrangian and Eulerian frames, for the direct-cascade regime in
two-dimensional fluid turbulence with air-drag-induced friction. We show
that different ways of extracting time scales from these time-dependent
structure functions lead to different dynamic-multiscaling exponents,
which are related to equal-time multiscaling exponents by different
classes of bridge relations; for a representative value of the friction
we verify that, given our error bars, these bridge relations hold. 
\end{abstract}
\maketitle

The scaling properties of both equal-time and time-dependent correlation
functions close to a critical point, say in a spin system, have been
understood well for nearly four 
decades~\cite{*[{See, e.g.,}][{and references therein}]{Cha+Lub98,hoh+hal77}}. 
By contrast, the development of a similar understanding of the multiscaling 
properties of equal-time and time-dependent structure functions in the inertial range
in fluid turbulence still remains a major challenge for it requires
interdisciplinary studies that must use ideas both from
nonequilibrium statistical mechanics and
turbulence~\cite{kol41,Fri96,lvo+pod+pro97,bif+bof+cel+tos99,kan+ish+got99,mit+pan03,
mit+pan04,mit+pan05,ray+mit+pan08,pan+ray+mit08}. We develop
here a complete characterization of the rich multiscaling
properties of time-dependent vorticity structure functions for the
direct-cascade regime of two-dimensional (2D) turbulence in fluid films
with friction, which we study via a direct numerical simulation (DNS).
Such a characterization has not been possible hitherto because it
requires very long temporal averaging to obtain good statistics for {\it
quasi-Lagrangian} structure functions~\cite{bel+lvo87}, which are
considerably more complicated than their conventional, Eulerian
counterparts as we show below.  Our DNS study yields a variety of
interesting results that we summarize informally before providing
technical details and precise definitions: (a) We calculate equal-time
and time-dependent vorticity structure functions in Eulerian and
quasi-Lagrangian frames~\cite{bel+lvo87}. (b) We then show how to extract an infinite
number of different time scales from such time-dependent structure
functions. (c) Next we present generalizations of the dynamic-scaling
Ansatz, first used in the context of critical phenomena~\cite{hoh+hal77} to
relate a diverging relaxation time $\tau$ to a diverging correlation
length $\xi$ via $\tau \sim \xi^z$, where $z$ is the dynamic-scaling
exponent. These generalizations yield, in turn, an infinity of
dynamic-multiscaling exponents~\cite{lvo+pod+pro97,bif+bof+cel+tos99,mit+pan03,
mit+pan04,mit+pan05,ray+mit+pan08,pan+ray+mit08}. (d)
A suitable extension of the multifractal formalism~\cite{Fri96}, which
provides a rationalization of the multiscaling of equal-time structure
functions in turbulence, yields linear bridge relations between
dynamic-multiscaling exponents and their equal-time
counterparts~\cite{lvo+pod+pro97,bif+bof+cel+tos99,mit+pan03,
mit+pan04,mit+pan05,ray+mit+pan08,pan+ray+mit08}; our study provides
numerical evidence in support of such bridge relations.  

The statistical properties of fully developed, homogeneous, isotropic
turbulence are characterized, {\it inter alia}, by the equal-time,
order-$p$, longitudinal-velocity structure function ${\mathcal S}_p(r)
\equiv \langle [\delta u_{\parallel}(r,t)]^p \rangle$, where $\delta
u_{\parallel}(r,t) \equiv [{\bf u}({\bf x} + {\bf r},t) - {\bf u}({\bf
x},t)\cdot {\bf r}/r]$,  ${\bf u}({\bf x},t)$ is the Eulerian velocity
at point ${\bf x}$ and time $t$, and $r\equiv \mid{\bf r}\mid$.  In the
inertial range $\eta_d \ll r \ll L$, ${\mathcal S}_p(r) \sim
r^{\zeta_p}$, where $\zeta_p$, $\eta_d$, and $L$, are, respectively, the
equal-time exponent, the dissipation scale, and the forcing scale. The
pioneering work~\cite{kol41} of Kolmogorov (K41) predicts simple scaling
with $\zeta_p^{K41} = p/3$ for three-dimensional (3D) homogeneous,
isotropic fluid turbulence. However, experiments and numerical
simulations show marked deviations from K41 scaling, especially for $p
\geq 4$, with $\zeta_p$ a nonlinear, convex function of $p$; thus, we
have multiscaling of equal-time velocity structure functions. To
examine dynamic multiscaling, we must obtain the order-$p$,
time-dependent structure functions ${\mathcal F}_p(r,t)$, which we
define precisely below, extract from these the time scales $ \tau_p(r)$,
and thence the dynamic-multiscaling exponents $z_p$ via 
dynamic-multiscaling Ans\"atze like $\tau_p(r) \sim r^{z_p}$. This task is
considerably more complicated than its analog for the determination of
the equal-time multiscaling exponents
$\zeta_p$ \cite{lvo+pod+pro97,bif+bof+cel+tos99,kan+ish+got99,mit+pan03,
mit+pan04,mit+pan05,ray+mit+pan08,pan+ray+mit08} for the following
two reasons: (I) In the conventional Eulerian description, the sweeping
effect, whereby large eddies drive all smaller ones directly, relates
spatial separations $r$ and temporal separations $t$ linearly via the
mean-flow velocity, whence we get trivial dynamic scaling with $z_p =1$,
for all $p$. A quasi-Lagrangian description~\cite{lvo+pod+pro97,bel+lvo87} eliminates 
sweeping effects so we calculate time-dependent, quasi-Lagrangian vorticity 
structure functions from our
DNS. (II) Such time-dependent structure functions, even for a fixed
order $p$, do not collapse onto a scaling function, with a unique,
order-$p$, dynamic exponent. Hence, even for a fixed order $p$, there is an
infinity of dynamic-multiscaling exponents~\cite{lvo+pod+pro97,bif+bof+cel+tos99,
mit+pan03,mit+pan04,mit+pan05,ray+mit+pan08,pan+ray+mit08}; roughly speaking, to
specify the dynamics of an eddy of a given length scale, we require this
infinity of exponents. 
 
Statistically steady fluid turbulence is very different in 3D and 2D;
the former exhibits a direct cascade of energy whereas the latter shows
an inverse cascade of kinetic energy from the energy-injection
scale to larger length scales and a direct cascade in which
the enstrophy goes towards small length scales~\cite{kraic67,*lei68,*bat69}; in many
physical realizations of 2D turbulence, there is an air-drag-induced
friction.  In this direct-cascade regime, velocity structure functions
show simple scaling but their vorticity counterparts exhibit
multiscaling~\cite{bof+cel+mus+ver02,tsa+ott+ant+guz05,per+pan09}, with exponents 
that depend on the friction.  
Time-dependent structure functions have not been studied in
2D fluid turbulence; the elucidation of the dynamic multiscaling of
these structure functions, which we present here, is an important step
in the systematization of such multiscaling in turbulence.

We numerically solve the forced, incompressible, 2D Navier-Stokes (2DNS) equation with
air-drag-induced friction, in the vorticity($\omega)$--stream-function($\psi)$ 
representation with periodic boundary conditions:
\begin{equation}
\partial_t \omega - J(\psi,\omega) = \nu \nabla^2 \omega - 
\mu \omega + f,
\label{eq:2DNS}
\end{equation}
where $\nabla^2 \psi = \omega$, 
$J(\psi,\omega) \equiv (\partial_x \psi)(\partial_y \omega) 
                  - (\partial_x \omega) (\partial_y \psi)$,
and the velocity
${\bf u } \equiv (-\partial_y \psi,  \partial_x \psi)$.
The coefficient of friction is $\mu$ and $f$ is the external force.
We work with both Eulerian and quasi-Lagrangian fields.  
The latter are defined with respect to a Lagrangian particle, 
which was at the point $\xizero$ at time $t_0$, and is at the position
$\xxi(t|\xizero,t_0)$ at time $t$, such that
$d\xxi(t|\xizero,t_0)/dt = {\bf u}[\xxi(t|\xizero,t_0),t]$, 
where ${\bf u}$ is the Eulerian velocity. The quasi-Lagrangian velocity field 
${\bf u}^{\rm QL}$ is defined~\cite{bel+lvo87} as follows:
\begin{equation}
{\bf u}^{\rm QL}({\bf x},t|\xizero,t_0) 
\equiv {\bf u}[{\bf x}+ \xxi(t|\xizero,t_0),t]; 
\label{eq:qldef}
\end{equation}
likewise, we can define the quasi-Lagrangian vorticity field $\omega^{\rm
QL}$ in terms of the Eulerian $\omega$. 
To obtain this quasi-Lagrangian
field we use an algorithm developed in Ref.~\cite{Mit05},
described briefly in the Supplementary Material.

\begin{figure*}
\begin{center}
\includegraphics[width=0.45\linewidth]{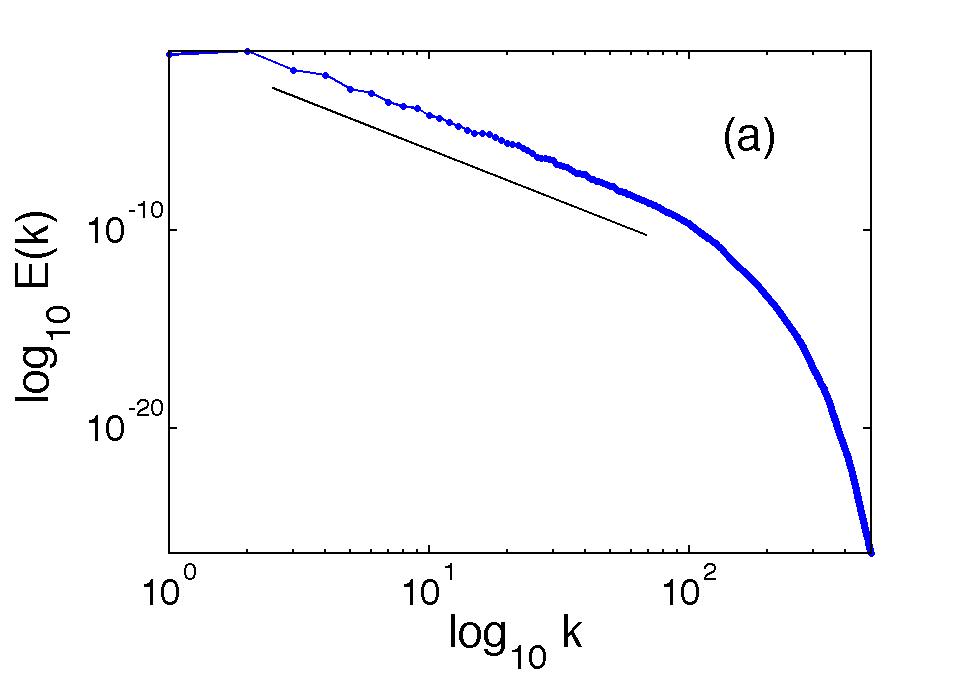}
\includegraphics[width=0.45\linewidth]{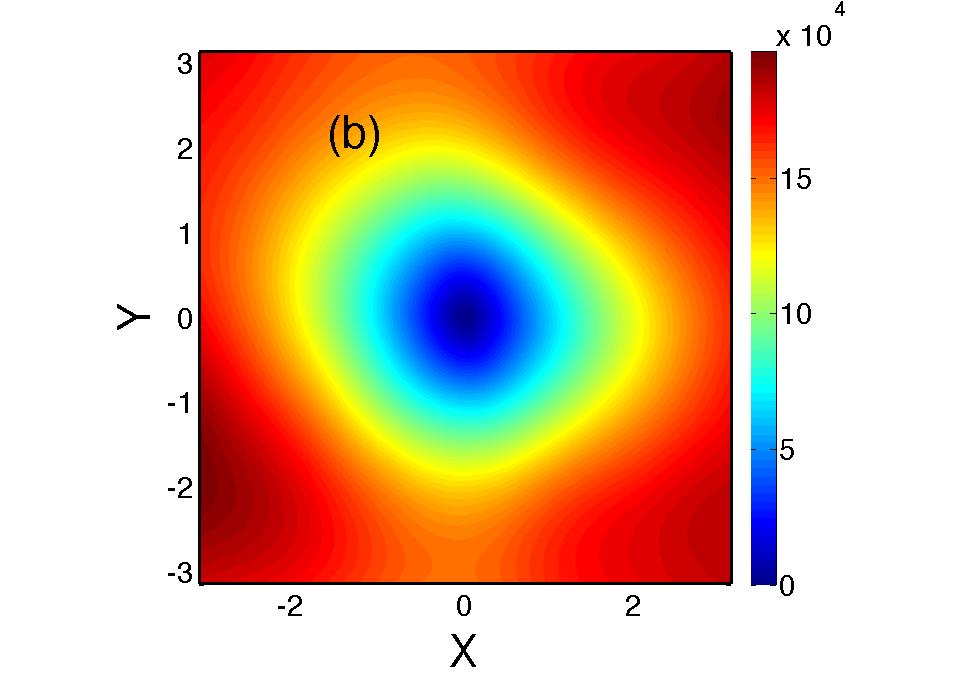} \\ 
\includegraphics[width=0.45\linewidth]{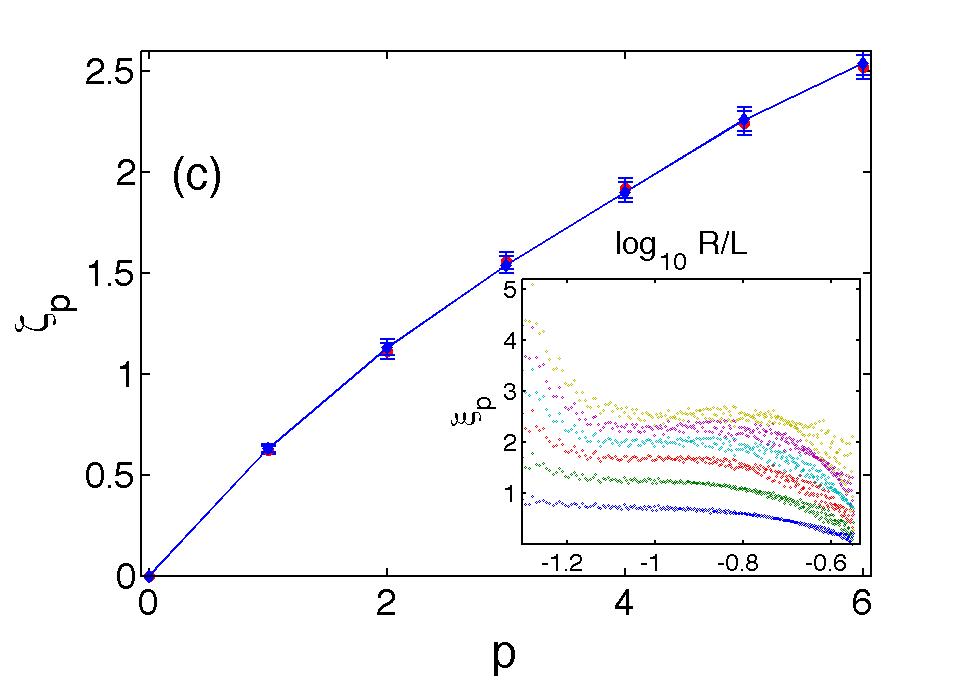}
\includegraphics[width=0.45\linewidth]{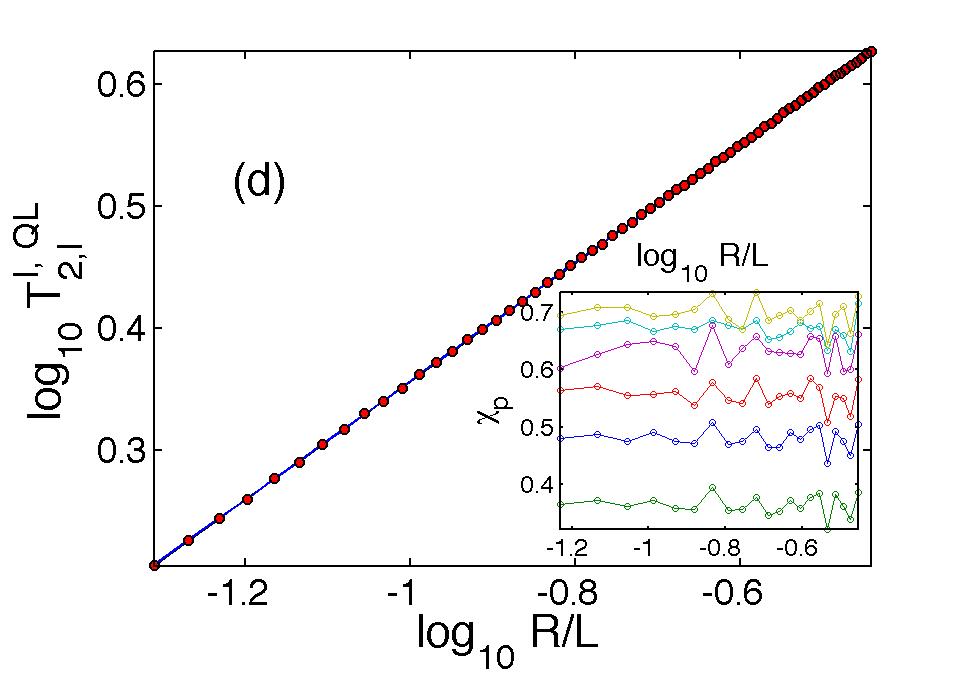}
\caption{(Color online) 
(a) Log-log (base 10) plot of the energy spectrum $E(k)$ versus $k$.
The black line with slope $-4.1$ is shown for reference. 
(b) Pseudocolor plot of the equal-time,
quasi-Lagrangian, second-order vorticity structure function  $S_2^{\rm
QL}({\bf R})$ averaged over the origin ${\bf r_c}$ (we use ${\bf r_c} =
(i,j), 2 \leq i, j \leq 5$); (c) plots of the equal-time, vorticity,
multiscaling exponents $\zeta^\phi_p$ versus $p$ for Eulerian (red circles) and
quasi-Lagrangian (blue diamonds) fields (error bars are comparable to the size of
the symbols); the inset shows the local slopes $\xi_p$, obtained as defined in the text, 
versus the separation, from $p$ = 1 (bottom) to $p$ = 6 (top); (d)  log-log
(base 10) plot of the order-2, degree-1, integral time scale $T^{I,{\rm
QL}}_{2,1}(R)$ versus the separation $R$ showing our data points (open
red circles) and the best-fit line (full black) in the scaling range;
the inset shows the local
slopes $\chi_p$, obtained as defined in the text, versus the separation, from 
$p$ = 1 (bottom) to $p$ = 6 (top).
}
\label{fig1}
\end{center}
\end{figure*}

To integrate the Navier-Stokes equations we use a pseudo-spectral method
with the $2/3$ rule for the removal of aliasing errors~\cite{per+pan09} and a
second-order Runge-Kutta scheme for time marching with a time step 
$\delta t = 10^{-3}$.  
We force the fluid deterministically on the second shell in Fourier space.
And we use $\mu = 0.1$, $\nu = 10^{-5}$, and $N = 2048^2$ collocation
points~\footnote{We have checked that $N = 1024^2$ collocation points
yield exponents that are consistent with those presented here. See S. S.
Ray, PhD Thesis, Indian Institute of Science, Bangalore (2010),
unpublished.} 
We obtain a turbulent but statistically steady
state with a Taylor microscale $\lambda \simeq 0.2$, Taylor-microscale
Reynolds number $Re_{\lambda} \simeq 1400$, and a box-size
eddy-turn-over time $\tau_{{\rm eddy}} \simeq 8$. 
We remove the effects of transients by discarding data upto
time $\lesssim 80 \tau_{{\rm eddy}}$. 
We then obtain data for averages of time-dependent structure functions for a duration of 
time $\simeq 100 \tau_{{\rm eddy}}$.
The energy spectrum  averaged over the same time interval is shown in 
Fig. (\ref{fig1}a).

The equal-time, order-$p$, vorticity structure functions we consider are
${\mathcal S}^\phi_p(r) \equiv \langle [\delta \omega^\phi (r,t)]^p
\rangle \sim r^{\zeta^\phi_p} $, for $\eta_d \ll r \ll L$, where $\delta
\omega^\phi (r,t) = [{\omega^\phi }({\bf x} + {\bf r} ,t) - {
\omega^\phi }({\bf x} ,t )]$, the angular brackets denote an average
over the nonequilibrium statistically steady state of the turbulent
fluid, and the superscript $\phi$ is either ${\rm E}$, in the Eulerian
case, or ${\rm QL}$, in the quasi-Lagrangian case; for notational
convenience we do not include a subscript $\omega$ on $S_p^\phi$ and the 
multiscaling exponent $\zeta^\phi_p$. We
assume isotropy here, but show below how to extract the isotropic parts
of $S_p^\phi$ in a DNS.  We also use the time-dependent,
order-$p$ vorticity structure functions
\begin{eqnarray}
{\mathcal F}^\phi_p(r,\{t_1,\ldots,t_p\}) \equiv 
        \langle [\delta \omega^{\phi } (r,t_1) \ldots  
                 \delta \omega^{\phi} (r,t_p)] \rangle;
\label{dynsp}
\end{eqnarray}
here $t_1, \ldots, t_p$ are $p$ different
times; clearly, ${\mathcal F}^\phi_p(r,\{t_1= \ldots =t_p=0\})={\mathcal
S}^\phi_p(r)$. We concentrate on the case $t_1=t_2=\ldots = t_l \equiv
t$ and $t_{l+1}=t_{l+2} = \ldots = t_p = 0$, with $l < p$,  and, for
simplicity, denote the resulting time-dependent structure function as
${\mathcal F}^\phi_p(r,t)$; shell-model studies~\cite{mit+pan03,mit+pan04} have
shown that the index $l$ does not affect dynamic-multiscaling exponents,
so we suppress it henceforth.  Given ${\mathcal F}^\phi_p(r,t)$,  it is
possible to extract a characteristic time scale $\tau_p(r)$ in many
different ways.  These time scales can, in turn, be used to extract the
order-$p$ dynamic-multiscaling exponents $z_p$ via the
dynamic-multiscaling Ansatz $ \tau_p(r) \sim r^{z_p}$.  If we obtain the
order-$p$, degree-$M$, {\it integral} time scale 
\begin{eqnarray}
 {\cal T}^{I,\phi}_{p,M}(r) \equiv 
 \biggl[ \frac{1}{{\mathcal S}^\phi_p(r)}
\int_0^{\infty}{\mathcal F}^\phi_p(r,t)t^{(M-1)} dt
\biggl]^{(1/M)},
\label{timp} 
\end{eqnarray} 
we can use it to extract the {\it integral} dynamic-multiscaling exponent
$z^{I,\phi}_{p,M}$ from the relation ${\cal T}^{I,\phi}_{p,M} \sim
r^{z^{I,\phi}_{p,M}}$.  Similarly, from the order-$p$, degree-$M$, {\it
derivative} time scale  
\begin{eqnarray}
 {\cal T}^{D,\phi}_{p,M} \equiv \biggl[\frac{1}{{\mathcal S}^\phi_p(r)}
                   \frac{\partial^M}{\partial t^M} 
                  {\mathcal F}^\phi_p(r,t) \biggl|_{t=0} \biggl]^{(-1/M)}, 
\label{tdpm}
\end{eqnarray} 
we obtain the {\it derivative} dynamic-multiscaling exponent
$z^{D,\phi}_{p,M}$ via the relation ${\cal T}^{D,\phi}_{p,M} \sim
r^{z^{D,\phi}_{p,M}}$. 

Equal-time vorticity structure functions in 2D fluid turbulence with
friction exhibit multiscaling in the direct cascade range~\cite{bof+cel+mus+ver02,
tsa+ott+ant+guz05,per+pan09}. 
For the case of 3D homogeneous, isotropic fluid turbulence, a generalization of the
multifractal model~\cite{Fri96}, which includes time-dependent velocity
structure functions~\cite{lvo+pod+pro97,mit+pan04,ray+mit+pan08,pan+ray+mit08}, yields linear bridge
relations between the dynamic-multiscaling exponents and their
equal-time counterparts.  For the direct-cascade regime in our study,
we replace velocity structure functions by vorticity structure
functions and thus obtain the following bridge relations 
for time-dependent vorticity structure functions in 2D fluid turbulence with friction:
\begin{equation}
z^{I,\phi}_{p,M} = 1 + [\zeta^\phi_{p-M} - \zeta^\phi_p]/M ;
\label{zipm}
\end{equation}
\begin{equation}
z^{D,\phi}_{p,M} = 1 + [\zeta^\phi_p - \zeta^\phi_{p+M}]/M.
\label{zdpm}
\end{equation}

The vorticity field $\omega^\phi = \langle \omega^\phi \rangle +
\omega^{\prime \phi}$ can be decomposed into the time-averaged mean flow
$ \langle \omega^\phi \rangle $ and the fluctuations $\omega^{\prime
\phi}$ about it.  To obtain good statistics for vorticity structure
functions it is important to eliminate any anisotropy in the flow by
subtracting out the mean flow from the field.  Therefore, we redefine
the order-$p$, equal-time structure function to be 
$S_p^\phi({\bf r_c},{\bf R}) \equiv \langle|\omega^{\prime \phi}({\bf r_c}+
       {\bf R})-\omega^{\prime \phi}({\bf r_c})|^p\rangle$, 
where ${\bf R}$ has magnitude $R$ and ${\bf r_c}$ is an origin. 
We next use  
$S_p^\phi({\bf R})\equiv \langle S_p^{\phi} ({\bf r_c},{\bf R}) 
        \rangle_{{\bf r_c}}$,
where the subscript ${\bf r_c}$ denotes an average over the origin (we
use ${\bf r_c} = (i,j), 2 \leq i, j \leq 5$). These averaged structure
functions are isotropic, to a good approximation for small $R$, as can
be seen from the illustrative pseudocolor plot of $S_2^{\rm QL}({\bf
R})$ in Fig.  (\ref{fig1}a).  The purely isotropic parts of such
structure functions can be obtained~\cite{per+pan09,bou+pro+sel05} via an
integration over the angle $\theta$ that ${\bf R}$ makes with the $x$
axis, i.e., we calculate $S_p^\phi(R)\equiv \int_{0}^{2\pi}
S_p^\phi({\bf R}) d\theta$ and thence the equal-time multiscaling
exponent $\zeta^\phi_p$, the slopes of the scaling ranges of log-log
plots of $S_p^\phi(R)$ versus $R$. The mean of the local slopes $\xi_p
\equiv d(\log S_p^\phi)/d(\log R)$ in the scaling range yields the
equal-time exponents; and their standard deviations give the error bars.
The equal-time vorticity multiscaling exponents, with $1 \le p \le 6$,
are given for Eulerian and quasi-Lagrangian cases in columns 2 and 3,
respectively, of Table 1; they are equal, within error bars, as can be
seen most easily from their plots versus $p$ in Fig.(\ref{fig1}c). 

We obtain the isotropic part of  ${\mathcal F}^\phi_p(R,t)$ in a
similar manner. Equations (\ref{timp}) and (\ref{tdpm}) now yield the
order-$p$, degree-$M$ integral and derivative time scales (see the
Supplementary Material).  Slopes of linear scaling ranges of log-log
plots of ${\cal T}^{I,{\phi}}_{p,M}(R)$ versus $R$ yield the dynamic
multiscaling exponent $z_{p,M}^{I,{\phi}}$. A representative plot for
the quasi-Lagrangian case, $p=2$ and $M=1$, is given in 
Fig. (\ref{fig1} d); 
we  fit over the range $-1.2 < \log_{10}(r/L)  < -0.55$ and obtain
the local slopes $\chi_p$ with successive, nonoverlapping sets of 3
points each. The mean values of these slopes yield our
dynamic-multiscaling exponents (column 5 in Table 1) and their standard
deviations yield the error bars.  We calculate the degree-$M$, order-$p$
derivative time exponents by using a sixth-order, finite-difference
scheme to obtain ${\cal T}^{D,{\phi}}_{p,M}$ and thence the
dynamic-multiscaling exponents $z_{p,M}^{D,{\phi}}$. Our results for the
quasi-Lagrangian case with $M=2$ are given in column 7 of Table 1. We
find, furthermore, that both the integral and derivative bridge
relations (\ref{zipm}) and (\ref{zdpm}) hold within our
error bars, as shown for the representative values of $p$ and $M$
considered in Table 1 (compare columns 4 and 5 for the integral relation
and columns 6 and 7 for the derivative relation). Note also that the
values of the integral and the derivative dynamic-multiscaling exponents
are markedly different from each other (compare columns 5 and 7 of Table
1).

The Eulerian structure functions ${\mathcal F}^{E}_p(R,t)$ also lead to nontrivial
dynamic-multiscaling exponents, which are equal to their
quasi-Lagrangian counterparts (see Supplementary Material).
The reason for this initially surprising result is that, in 2D turbulence,
the friction controls the size of the largest vortices, provides 
an infra-red cut-off
 at large length scales, and thus suppresses the sweeping effect.
We have demonstrated this in the supplementary material.
Had the sweeping effect not been suppressed, we would have obtained
trivial dynamic scaling for the Eulerian case.

The calculation of dynamic-multiscaling exponents has been limited so
far to shell models for 3D, homogeneous, isotropic 
fluid \cite{bif+bof+cel+tos99,mit+pan03,mit+pan04,ray+mit+pan08,pan+ray+mit08} 
and passive-scalar turbulence~\cite{mit+pan05}. We have
presented the first study of such dynamic multiscaling in the
direct-cascade regime of 2D fluid turbulence with friction by
calculating both quasi-Lagrangian and Eulerian structure functions. Our
work brings out clearly the need for an infinity of time scales and
associated exponents to characterize such multiscaling; and it verifies,
within the accuracy of our numerical calculations, the linear bridge
relations (\ref{zipm}) and (\ref{zdpm}) for a representative value of
$\mu$. We find that friction also suppresses
sweeping effects so, with such friction, even Eulerian vorticity
structure functions exhibit dynamic multiscaling with exponents that are
consistent with their quasi-Lagrangian counterparts.
 
Experimental studies of Lagrangian quantities in turbulence have been
increasing steadily over the past decade~\cite{ott+man00,*por+vot+cra+ale+bod02,
*mor+met+mic+pin01}. We hope, therefore,
that our work will encourage studies of dynamic multiscaling in 
turbulence.  Furthermore, it will be interesting to check whether the
time scales considered here can be related to the persistence time
scales for 2D turbulence~\cite{per+ray+mit+pan11}.

\begin{table*}
\framebox{\begin{tabular}{c|c|c|c|c|c|c}
order$(p)$ & $\zeta^{{\rm E}}_p$ &$\zeta^{{\rm QL}}_p$ & $z^{I,{\rm QL}}_{p,1}$[Eq.(\ref{zipm})] & $z^{I,{\rm QL}}_{p,1}$
 & $z^{D,{\rm QL}}_{p,2}$[Eq.(\ref{zdpm})] & $z^{D,{\rm QL}}_{p,2}$  \\
\hline
 1 & 0.62 $\pm$ 0.009 &  0.63 $\pm$ 0.008 & 0.366 $\pm$ 0.008  & 0.37 $\pm$ 0.02
    & 0.55  $\pm$ 0.02  & 0.53 $\pm$ 0.02
   \\
 2 &  1.13 $\pm$ 0.009 & 1.13 $\pm$ 0.008  & 0.50 $\pm$ 0.02 & 0.48 $\pm$ 0.01
   & 0.62 $\pm$ 0.02  & 0.62  $\pm$  0.2
 \\
 3 & 1.561 $\pm$ 0.009 & 1.54  $\pm$ 0.01  & 0.59 $\pm$ 0.02 & 0.57 $\pm$ 0.01
   & 0.64 $\pm$ 0.02  & 0.65 $\pm$ 0.01
  \\
 4 & 1.92  $\pm$ 0.01  & 1.90  $\pm$ 0.01  &  0.64 $\pm$ 0.02 & 0.63 $\pm$ 0.01
 & 0.68 $\pm$ 0.03  & 0.68  $\pm$ 0.01  \\

 5 & 2.24 $\pm$ 0.01 & 2.26 $\pm$ 0.01 & 0.64 $\pm$ 0.02  & 0.65 $\pm$ 0.02
   & 0.70  $\pm$ 0.03  & 0.70 $\pm$ 0.02
  \\
 6 &  2.52 $\pm$ 0.02 &  2.54 $\pm$ 0.02 & 0.72  $\pm$ 0.03  & 0.67 $\pm$ 0.02
   & 0.71 $\pm$ 0.04  & 0.71 $\pm$ 0.03 
   \\

\end{tabular}}
\caption{ Order-$p$ (column 1); equal-time,  
Eulerian exponents $\zeta^{\rm E}_p$
(column 2); equal-time, quasi-Lagrangian exponents $\zeta^{\rm QL}_p$
(column 3); integral-scale, dynamic-multiscaling exponent
$z^{I,{\rm QL}}_{p,1}$ (column 4) from the bridge relation 
and the values of $\zeta^{\rm QL}_p$ in column 3; $z^{I,{\rm QL}}_{p,1}$ 
from our
calculations of time-dependent structure functions (column 5);
the derivative-time exponents
$z^{D,{\rm QL}}_{p,2}$ (column 6) from the bridge relation and the values of
$\zeta^{\rm QL}_p$ in column 3; $z^{D,{\rm QL}}_{p,2}$ from our calculations
of the time-dependent structure function (column 7).
The error estimates are obtained as described in the text.}
\end{table*}
\begin{acknowledgments}
We thank J. K. Bhattacharjee for discussions, the European Research
Council under the Astro-Dyn Research Project No. 227952, 
National Science Foundation under Grant No. PHY05-51164,
CSIR, UGC, and DST (India) for support, and SERC (IISc) for computational 
resources. PP and RP are members of the International Collaboration for Turbulence
Research; RP, PP, and SSR acknowledge support from the COST Action
MP0806.  Just as we were preparing this study for publication we became
aware of a recent preprint~\cite{bif+cal+tos11} on a related study for
3D fluid turbulence. We thank L. Biferale for sharing
the preprint of this paper with us.
\end{acknowledgments}

\appendix
\section{Algorithm for obtaining quasi-Lagrangian fields in a 
pseudospectral simulation}

To obtain a quasi-Lagrangian field from its Eulerian counterpart, we
track a single Lagrangian particle by using a bilinear-interpolation
method~[17].  If we replace the Eulerian velocity in
Eq.~(2) by its Fourier-integral representation, we obtain
\begin{equation}
{\bf u}^{\rm QL}({\bf x},t | \xizero,t_0) =
    \int {\bf {\hat u}}({\bf q},t)
  \exp[i {\bf q}\cdot ( {\bf x} +\xxi(t|\xizero,t_0),t) ) ]  d{\bf q},
\nonumber
\end{equation}
where ${\bf q}$ is the wave vector.  In the pseudospectral algorithm we
use to solve Eq.~(1),
the quasi-Lagrangian velocity is defined with respect to a Lagrangian particle, 
which was at the point $\xizero$ at time $t_0$, and is at the position
$\xxi(t|\xizero,t_0)$ at time $t$, such that
$d\xxi(t|\xizero,t_0)/dt = {\bf u}[\xxi(t|\xizero,t_0),t]$. 
We calculate ${\bf {\hat u}}({\bf q},t)$; thus, the Fourier integral 
above can be evaluated at each time
step by an additional call to a fast-Fourier-Transform (FFT) subroutine.
The additional computational cost of obtaining ${\bf u}^{\rm QL}$ at all
collocation points is that of following a single Lagrangian particle and
an additional FFT at each time step.

\section{Numerical determination of integral time scales from
time-dependent structure functions}

To extract the integral time scale, of degree $M$, from a time-dependent
structure function, we have to evaluate the integral in Eq.~(4)
numerically.  In practice, because of poor statistics at long times, we
integrate from $t = 0$ to $t = t_*$, where $t_*$ is the time at which
${\mathcal F}^{\phi}_p(R',t) = \epsilon$; we choose $\epsilon = 0.6$,
but we have checked that our results do not change, within our error
bars, for $0.5 \le \epsilon \le 0.75$. This numerical integration is 
done by using the trapezoidal rule.

\section{Dynamic multiscaling for Eulerian structure functions}

Equal-time Eulerian structure functions have been discussed in our paper
above.  To obtain time-dependent, Eulerian, vorticity structure
functions we proceed as we did in the quasi-Lagrangian case. We obtain
the required vorticity increments and from these the purely isotropic
part of the time-dependent, order-$p$ structure function ${\mathcal
F}^{\rm E}_p(R',t)$.  Equations (4) and (5) now yield
the order-$p$, degree-$M$ integral and derivative Eulerian time scales.
For the former we should integrate ${\mathcal F}^{\rm E}_p(R',t)$ from
$t=0$ to $t = \infty$; in practice, because of poor statistics at long
times, we integrate from $t = 0$ to $t = t_*$, where $t_*$ is the time
at which ${\mathcal F}^{\rm E}_p(R',t) = \epsilon$; we choose $\epsilon
= 0.6$, but we have checked that our results do not change, within our
error bars, for $0.5 \le \epsilon \le 0.75$. Slopes of linear scaling
ranges of log-log plots of ${\cal T}^{I,{\rm E}}_{p,M}(R')$ versus $R'$
yield the dynamic multiscaling exponent $z_{p,1}^{I,{\rm E}}$. A
representative plot for the Eulerian case, $p=2$, and $M=1$ is given in
Fig. (\ref{figsupp1} a); we  fit over the range $-1.2 < \log_{10}(r/L)
< -0.55$ and obtain the local slopes $\chi_p$ with successive,
non-overlapping sets of 3 points each. The mean values of these
slopes yield our dynamic-multiscaling exponents (column 4 in Table ~\ref{table_eu})
and their standard deviations yield the error bars.  We calculate the
degree-$M$, order-$p$ derivative time exponents by using a sixth-order
finite difference scheme to obtain ${\cal T}^{D,{\rm E}}_{p,M}$ and
thence the dynamic-multiscaling exponents $z_{p,M}^{D,{\rm E}}$; data
for the Eulerian case and the representative value $M=2$ are given in
column 6 of Table~\ref{table_eu}. We find, furthermore, that both the integral and
derivative bridge relations, Eq.~(6), and Eq.~(7).
hold within our error bars, as shown for the representative values of
$p$ and $M$ considered in Table \ref{table_eu} (compare columns 3 and 4 for the
integral relation and columns 5 and 6 for the derivative relation). The
values of the integral and the derivative dynamic-multiscaling exponents
are markedly different from each other (compare columns 4 and 6 of Table
\ref{table_eu}) and the plots of these exponents versus $p$ in Fig. (\ref{figsupp1}
b).  In Fig. (\ref{figsupp1} c), we make the same comparison for the
quasi-Lagrangian case.  Furthermore, a comparison of the
quasi-Lagrangian and Eulerian dynamic-multiscaling exponents given in
Tables I in the original paper and Table \ref{table_eu}, respectively, show that these 
are the same (within our error bars).

\section{Demonstration of Infra-red cutoff of the inverse cascade}
We have shown that in two dimensional turbulence with friction,
the Eulerian and the quasi--Lagrangian velocities have the same dynamical
exponents. This is because the inverse cascade has a friction--dependent 
infra-red cutoff. 

\begin{figure*}
\includegraphics[width=0.45\linewidth]{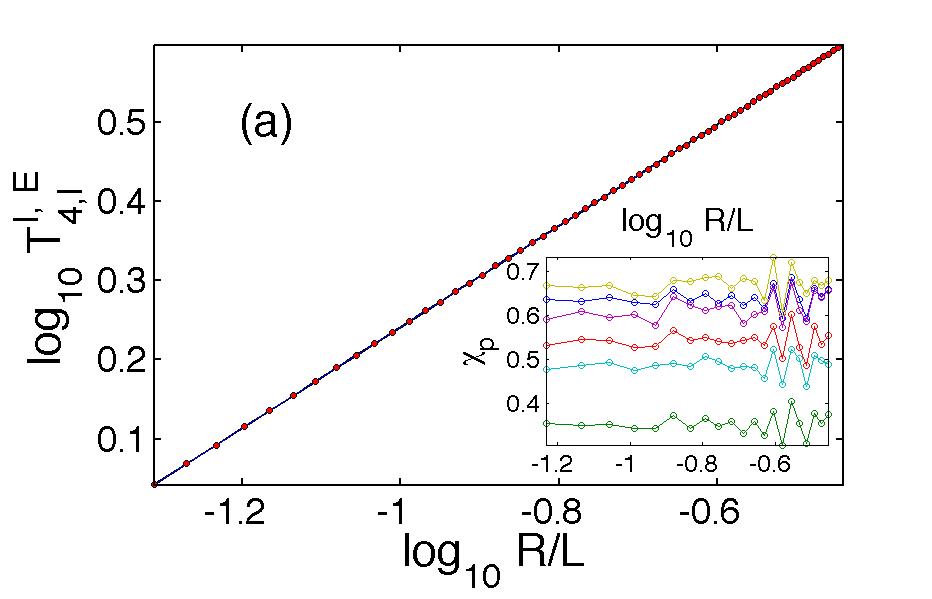} 
\includegraphics[width=0.45\linewidth]{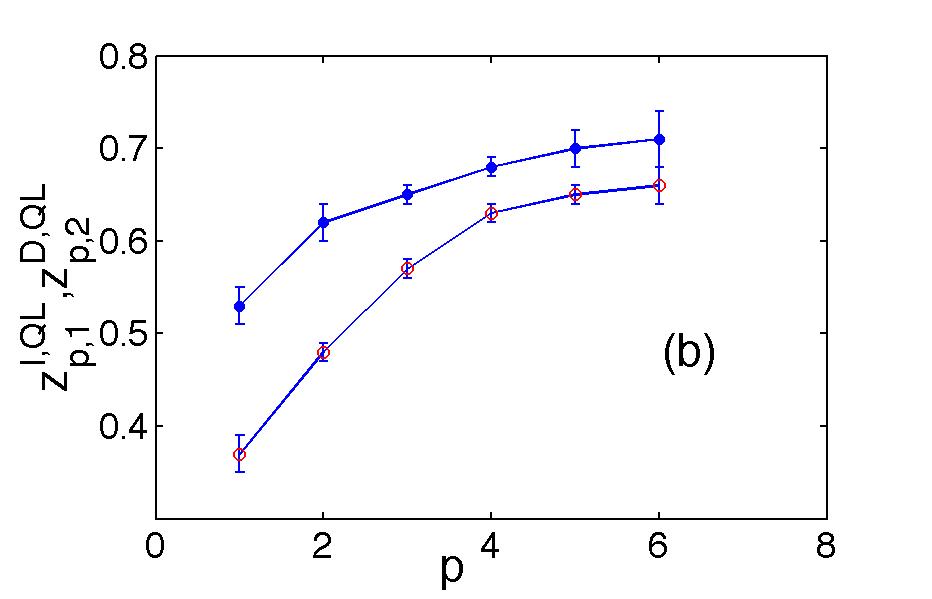} \\ 
\includegraphics[width=0.45\linewidth]{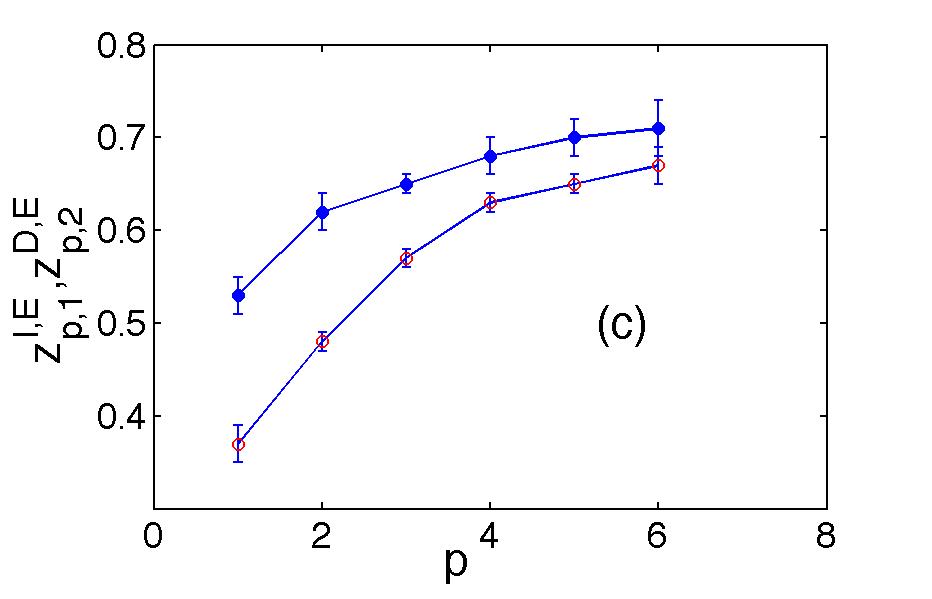}
\includegraphics[width=0.45\linewidth]{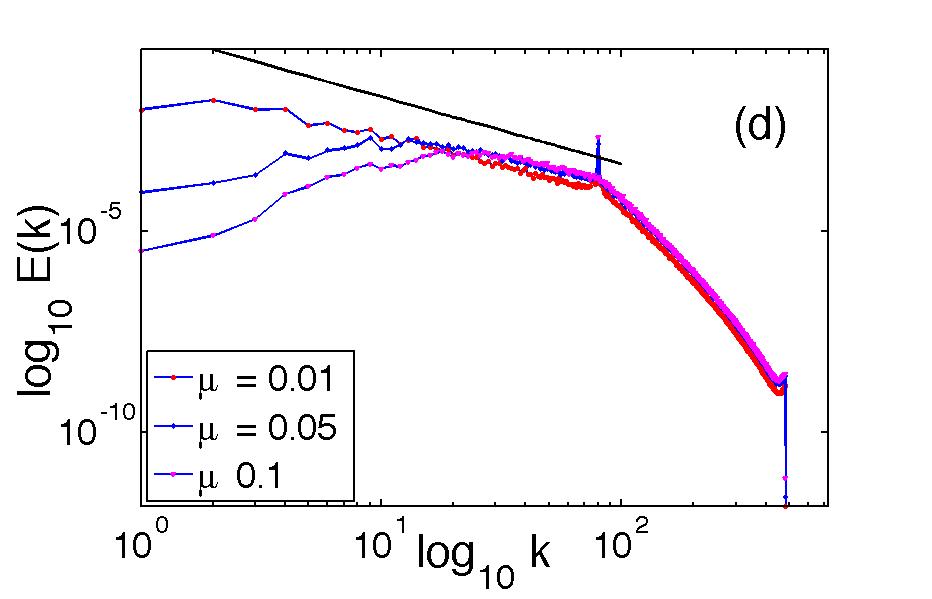}
\caption{(Color online) (a)  Log-log (base 10) plot of the order-2, degree-1, integral time scale 
$T^{I,{\rm E}}_{4,1}(R')$ versus the separation $R'$ showing our data points
(open red circles) and the best-fit line (full black) in the scaling
range; the inset shows the local slopes $\chi_p$, obtained as defined in the text,
versus the separation, from $p=1$ (bottom) tp $p = 6$ (top); 
(b) plots of the vorticity, dynamic-multiscaling, quasi-Lagrangian
exponents $z_{p,1}^{I,{\rm QL}}$ (open red circles) and $z_{p,2}^{D,{\rm
QL}}$ (full blue circles) versus $p$ with the error bars given in columns 4
and 6, respectively, in Table I;
(c) plots of the vorticity, dynamic-multiscaling, Eulerian
exponents $z_{p,1}^{I,{\rm E}}$ (open red circles) and $z_{p,2}^{D,{\rm
E}}$ (full blue circles) versus $p$ with the error bars given in columns 4
and 6, respectively, in Table II.
(d) log-log (base 10) plot of the energy spectrum $E(k)$ versus the
wave-vector magnitude $k$ for $\mu=0.01$ (red filled circles),
$\mu=0.05$ (blue filled diamonds), and $\mu=0.1$ (magenta filled
triangles); the peak is at the injection scale $k = 80$ and the black line
indicates the K41, 2D-inverse-cascade slope.}
\label{figsupp1}
\end{figure*}

To illustrate the
development of this cutoff scale,  we have carried out DNS studies of 2D
fluid turbulence with  $\mu = 0.01, \, 0.05$, and $0.1$, $1024^2$
collocation points, and forcing at a wave-vector magnitude $k=80$; our
DNS studies resolve the inverse-cascade regime in the statistically
steady state. The energy spectra from these DNS studies, plotted in  
Fig.~(\ref{figsupp1}d), show clearly that, as $\mu$ increases, the inverse
cascade is cut off at ever larger values of $k$.  Thus, the
friction produces a regularization of the flow and suppresses infrared
(sweeping) divergences.
\begin{table*}
\framebox{\begin{tabular}{c|c|c|c|c|c}
order$(p)$ & $\zeta^{{\rm E}}_p$ & $z^{I,{\rm E}}_{p,1}$[Eq.(6)] & $z^{I,{\rm E}}_{p,1}$
 & $z^{D,{\rm E}}_{p,2}$[Eq.(7)] & $z^{D,{\rm E}}_{p,2}$  \\
\hline
 1 &  0.62 $\pm$ 0.009 & 0.372 $\pm$ 0.009  & 0.37 $\pm$ 0.02
    & 0.53  $\pm$ 0.02  & 0.53 $\pm$ 0.02
   \\
 2 &  1.13 $\pm$ 0.009  & 0.51 $\pm$ 0.02 & 0.48 $\pm$ 0.01
   & 0.60 $\pm$ 0.02  & 0.62  $\pm$  0.02
 \\
 3 & 1.561  $\pm$ 0.009  & 0.56 $\pm$ 0.02 & 0.57 $\pm$ 0.01
   & 0.66 $\pm$ 0.02  & 0.65 $\pm$ 0.02
  \\
 4 & 1.92 $\pm$ 0.01  &  0.64 $\pm$ 0.02 & 0.63 $\pm$ 0.01
 & 0.70 $\pm$ 0.03  & 0.68  $\pm$ 0.02  \\

 5 & 2.24 $\pm$ 0.01 & 0.68 $\pm$ 0.02  & 0.65 $\pm$ 0.02
   & 0.71  $\pm$ 0.03  & 0.70 $\pm$ 0.02
  \\
 6 &  2.52 $\pm$ 0.02 & 0.72  $\pm$ 0.03  & 0.67 $\pm$ 0.02
   & 0.72 $\pm$ 0.03  & 0.71 $\pm$ 0.03 
   \\
\end{tabular}}
\caption{ Order-$p$ (column 1); equal-time,  
Eulerian exponents $\zeta^{\rm E}_p$
(column 2); integral-scale, dynamic-multiscaling exponent
$z^{I,{\rm E}}_{p,1}$ (column 3) from the bridge relation 
and the values of $\zeta^{\rm E}_p$ in column 2; $z^{I,{\rm E}}_{p,1}$ 
from our
calculation of time-dependent structure functions (column 4);
the derivative-time exponents
$z^{D,{\rm E}}_{p,2}$ (column 5) from the bridge relation and the values of
$\zeta^{\rm E}_p$ in column 2; $z^{D,{\rm E}}_{p,2}$ from our calculation
of time-dependent structure function (column 6).
The error estimates are obtained as described in the text.}
\label{table_eu}
\end{table*}

\begin{thebibliography}{10}

\bibitem{Cha+Lub98}
P. Chaikin and T. Lubensky, {\em Principles of condensed matter physics}
  (Cambridge, Cambridge University Press, UK, 1998).

\bibitem{hoh+hal77}
P.~C. {Hohenberg} and B.~I. {Halperin}, Rev. Mod. Phys. {\bf 49},  435  (1977).

\bibitem{kol41}
A. Kolmogorov, Dokl. Acad. Nauk USSR {\bf 30},  9  (1941).

\bibitem{Fri96}
U. Frisch, {\em Turbulence the legacy of A.N. Kolmogorov} (Cambridge University
  Press, Cambridge, 1996).

\bibitem{lvo+pod+pro97}
V. L'vov, E. Podivilov, and I. Procaccia, Phys. Rev. E {\bf 55},  7030  (1997).

\bibitem{bif+bof+cel+tos99}
L. Biferale, G. Bofetta, A. Celani, and F. Toschi, Physica D {\bf 127},  187
  (1999).

\bibitem{kan+ish+got99}
Y. Kaneda, T. Ishihara, and K. Gotoh, Phys. Fluids. {\bf 11},  2154  (1999).

\bibitem{mit+pan03}
D. Mitra and R. Pandit, Physica A {\bf 318},  179  (2003).

\bibitem{mit+pan04}
D. Mitra and R.Pandit, Phys. Rev. Lett {\bf 93},  024501  (2004).

\bibitem{mit+pan05}
D. Mitra and R.Pandit, Phys. Rev. Lett {\bf 95},  144501  (2005).

\bibitem{ray+mit+pan08}
S. Ray, D. Mitra, and R. Pandit, New. J. Phys {\bf 10},  033003  (2008).

\bibitem{pan+ray+mit08}
R. Pandit, S. Ray, and D. Mitra, Eur. Phys. J. B {\bf 64},  463  (2008).

\bibitem{bel+lvo87}
V. Belinicher and V. L'vov, Sov. Phys. JETP {\bf 66},  303  (1987).

\bibitem{kraic67}
R. Kraichnan, Physics of Fluids {\bf 10},  1417  (1967).

\bibitem{lei68}
C. Leith, Physics of Fluids {\bf 11},  671  (1968).

\bibitem{bat69}
G. Batchelor, Phys. Fluids {\bf 12},  II-233  (1969).

\bibitem{bof+cel+mus+ver02}
G. Boffetta, A. Celani, S. Musacchio, and M. Vergassola, Phys. Rev. E {\bf 66},
   026304  (2002).

\bibitem{tsa+ott+ant+guz05}
Y. {Tsang}, E. Ott, T. Antonsen, and P. Guzdar, Phys. Rev. E {\bf 71},  066313
  (2005).

\bibitem{per+pan09}
P. Perlekar and R. Pandit, New J. Phys. {\bf 11},  073003  (2009).

\bibitem{Mit05}
D. Mitra, Ph.D. thesis, Dept. of Physics, Indian Institute of Science,
  Bangalore., 2005.

\bibitem{Note1}
We have checked that $N = 1024^2$ collocation points yield exponents that are
  consistent with those presented here. See S. S. Ray, PhD Thesis, Indian
  Institute of Science, Bangalore (2010), unpublished.

\bibitem{bou+pro+sel05}
E. Bouchbinder, I. Procaccia, and S. Sela, Phys. Rev. Lett. {\bf 95},  255503
  (2005).

\bibitem{ott+man00}
S. Ott and J. Mann, J. Fluid Mech. {\bf 422},  207  (2000).

\bibitem{por+vot+cra+ale+bod02}
A.~L. Porta {\it et~al.}, Nature(London) {\bf 409},  1017  (2001).

\bibitem{mor+met+mic+pin01}
N. Mordant, P. Metz, O. Michel, and J.-F. Pinton, Phys. Rev. Lett. {\bf 87},
  214501  (2001).

\bibitem{per+ray+mit+pan11}
P. Perlekar, S. Ray, D. Mitra, and R. Pandit, Phys. Rev. Lett {\bf 106},
  054501  (2011).

\bibitem{bif+cal+tos11}
L. Biferale, E. Calzavarini, and F. Toschi, Phys. Fluids {\bf 23},  085107
  (2011).

\end{thebibliography}
\end{document}